\documentclass{aa}
\usepackage{graphicx}
\usepackage{txfonts}

\begin{document}

\title{V4140 Sgr: a short period dwarf nova with a peculiar
  behavior\thanks{Based on observations made at the Laborat\'{o}rio
    Nacional de Astrof\'{\i}sica (LNA/MCT), Brazil.}}

\author{B. W. Borges
        \and
        R. Baptista}

\offprints{B. W. Borges}

\institute{Departamento de F\'{\i}sica,
           Universidade Federal de Santa Catarina, CEP 88040-900,\\
           Florian\'{o}polis, Brazil\\
           \email{bernardo@astro.ufsc.br, bap@astro.ufsc.br}
          }

\date{}

\abstract{ We report on time series of CCD photometry of V4140 Sgr
  between 1991 and 2001. The analysis reveals that the object was in
  the decline from an outburst in 1992 and again in outburst in 2001.
  The historical light curve collected by amateur astronomers shows
  low amplitude ($1-2$ mag) outbursts $5-10$ days long, recurrent
  every 80-90 days, confirming its dwarf nova nature. We derive an
  outburst decline time scale of $1.2$ days\,mag$^{-1}$.  The orbital
  parameters are revised.  We find a mass ratio $q = M_{1}/M_{2} =
  0.125 \pm 0.015$, an inclination $i = 80.2^{\circ} \pm 0.5^{\circ}$,
  and $M_{1} = 0.73 \pm 0.08 M_{\odot}$ and $M_{2} = 0.09 \pm 0.02
  M_{\odot}$, respectively for the primary and secondary star masses.
  The predicted values for the semi-amplitude of the radial velocity
  curve of the primary and secondary stars are $K_{1} = 55 \pm 7$ km/s
  and $K_{2}=442 \pm 15$ km/s, respectively.  Eclipse mapping
  techniques were applied to data both in quiescence and in outburst
  to derive accretion disc surface brightness maps. A distance of $d =
  170 \pm 30$ pc is obtained from a method similar to that used to
  constrain the distance to open clusters.  From this distance, disc
  radial brightness temperature distributions are determined. The
  temperatures in the quiescent disc vary from $6\,000$ K in the inner
  regions to $3\,000$ K near the outer disc and are flatter than the
  $T \propto r^{-\frac{3}{4}}$ law for optically thick steady-state
  disc. The outburst occurs mainly with a significant increase in
  brightness of the intermediate and outer disc regions.  The disc
  temperatures remain below the critical effective temperature $T_{\rm
    crit}$ at all disc radii during outburst. The radial temperature
  distributions in quiescence and in outburst are significantly
  different from those of other dwarf novae of similar orbital period.
  These results cannot be explained within the framework of the disc
  instability model. We suggest that the small amplitude outbursts of
  V4140 Sgr are caused by bursts of enhanced mass transfer rate from
  the secondary star.

\keywords{accretion, accretion discs -- eclipses --
  stars:binaries:close -- stars:individual:V4140 Sgr -- novae,
  cataclysmic variables}}

\maketitle

\section{Introduction}
\label{sec:intro}

Dwarf novae are cataclysmic variable stars (CVs) which show recurrent
outbursts typically of 2 to 5 mag on time-scales of weeks to months,
caused by a sudden increase in the mass accretion rate. SU UMa stars
are a subclass of the dwarf novae which exhibit occasional
superoutbursts $\sim 0.7$ mag brighter and $\sim 3-5$ times longer
than the regular outbursts.

Two models were proposed back in the 70's to explain the cause of the
outbursts. In the mass transfer instability model (MTI), the outburst
are the response of a high viscosity accretion disc to a burst of
matter transferred from the donor star.  In the disc instability model
(DI), the outbursts are caused by a thermal-viscous instability cycle
in the disc, in which an annulus in a low viscosity state disc reaches
a critical condition which propagates as a heating wave and
progressively switches the disc to a high viscosity state, causing the
rapid inward diffusion of the orbiting gas (see Lasota 2001 for a
review). According to the DI model, there is a critical effective
temperature $T_{\rm crit}$, below which the disc gas should remain to
allow the thermal instability to set in, and above which the disc gas
should stay while in outburst (e.g. Warner 1995),

\begin{equation}
T_{\rm crit}(r) = 7476 \left( \frac{r}{R_{L_{1}}} \right)^{-0.105}
\left( \frac{M_{1}}{0.75 M_{\odot}} \right)^{-0.15}\ \ \ \ {\rm K,}
\label{eq:tcrit}
\end{equation}

\noindent
where $R_{L_{1}}$ is the distance from the disc centre to the inner
Lagrangian point and $M_{1}$ is the primary mass. 

Over the last two decades the DI model became the dominant theory to
explain dwarf novae outburst because it could reproduce observed
phenomena seemingly much better than the MTI model (e.g., Cannizzo
1993) and because there is a physical mechanism for the disc
instability, which has not been the case for the MTI model (Gontikakis
\& Hameury 1993, Warner 1995, Lasota 2001).  However, the present
situation is more complex.  Mass transfer variations became an
important element of the DI model.  Many aspects of dwarf novae
outburst can only be reasonably well reproduced by state-of-the-art DI
simulations with the inclusion of variations in mass transfer from the
secondary star (King \& Cannizzo 1998; Schreiber et al. 2000, 2003;
Buat-M\'{e}nard et al. 2001).  On the other hand, there are still
unsolved problems in the DI model (Smak 2000) and recent detailed
observations of some dwarf novae are in contradiction with the DI
model (Hellier et al. 2000; Baptista \& Catal\'{a}n 2001; Baptista \&
Bortoletto 2004).

V4140 Sgr is a relatively faint, eclipsing CV ($V \simeq 18$
mag) with a short orbital period ($P_{{\rm orb}} \simeq 88$ min). The
eclipses were discovered by Jablonski \& Steiner (1987). The
spectroscopy carried out by Mukai et al. (1988) confirmed its
classification as a CV and allowed the measurement of a systemic
velocity of $\gamma = 58 \pm 5$ km/s and a primary radial velocity
semi-amplitude of $K_{1} = 56 \pm 7$ km/s. Baptista et al. (1989, here
after BJS) derived the orbital parameters for the object from
high-speed photometry covering 25 eclipses. Baptista \& Steiner (1991)
applied eclipse mapping techniques to investigate the structure of the
accretion disc of V4140 Sgr. Because their data consisted of
uncalibrated white light curves, it was not possible to determine the
radial brightness distribution from the eclipse maps.  Also, because
their light curves were normalized to the out-of-eclipse level, they
were not able to distinguish observations in quiescence from those in
outburst. Downes et al. (2001) lists V4140 Sgr as an
SU UMa star, however until recently there was no observations
of outbursts or superoutbursts to confirm that.

This paper reports the analysis of time series $BVR$ CCD photometry of
V4140 Sgr. Its is organized as follows.
Section~\ref{sec:observ} describes the observations and the data
reduction procedures, and Section~\ref{sec:dwarfn} addresses the
observations which characterize the dwarf nova behavior of the star.
Section~\ref{sec:data_an} presents the data analysis: the revised
orbital parameters (\S\ref{sub:orbit}) and the eclipse mapping
analysis of the $BVR$ light curves (\S\ref{sub:emm}). The results are
presented in Section~\ref{sec:resul} and discussed in
Section~\ref{sec:disc}.  Section~\ref{sec:concl} summarizes the main
conclusions.

\section{The observations and data reduction}
\label{sec:observ}

The observations were performed with the 1.6~m Perkin-Elmer telescope
of Laborat\'{o}rio Nacional de Astrof\'{\i}sica (LNA/MCT - Brazil)
between 1991 and 2001, comprising a total of 39 eclipses. All data
were obtained with high-speed CCD photometry in the $BVR$ bands.
Observations before 1996 employed a EEV CCD array with $770 \times
1152$ pixels of $22.5 \times 22.5$ $\mu$m. The remaining runs were
obtained with a blue sensitive, back illuminated EEV CCD with $385
\times 578$ pixels of $22 \times 22$ $\mu$m. The plate scale is $\sim
0.6\,\arcsec$\,pixel$^{-1}$ for both detectors.
Table~\ref{tab:journal} shows the journal of the observations. $\Delta
t$ is the integration time and $N$ is the number of points of each
run. The last three columns list the phase range of the observed run,
the quality of the night sky and the range of seeing (in arcseconds).

\begin{table*}[ht]
\centering
\caption{Journal of the observations}  \label{tab:journal}
\small
\begin{tabular}{@{}cccclclcc@{}}
\hline\hline
Date (UT) & Start (HJD) & Cycle & Passband &\multicolumn{1}{c}{$\Delta
t$} & $N$ &\multicolumn{1}{c}{Phase} & Night & Seeing \\ 
& $(2400000+)$ & & &\multicolumn{1}{c}{(s)} &
&\multicolumn{1}{c}{Range} & Quality$^{\dagger}$ & (\arcsec) \\ [1ex] \hline
1991 Jul 08 & 48446.71492 & 35\,570 &$V$&  15,40   & 106 & $-$0.15,+0.5  & B & 1.5 -- 2.5\\
    "       & 48446.75568 & 35\,571 &$V$&  15,40   & 133 & $-$0.5,+0.45  & B & \\
    "       & 48446.81840 & 35\,572 &$V$&  15,40   & 115 & $-$0.47,+0.26 & B & \\
1991 Jul 09 & 48447.57448 & 35\,584 &$V$&   15     & 127 & $-$0.16,+0.5  & B & 1.5 -- 2.5\\
    "       & 48447.61545 & 35\,585 &$V$&   15     & 162 & $-$0.5,$-$0.44  & B & \\
    "       & 48447.70001 & 35\,586 &$V$&   15     & 118 & $-$0.12,+0.5  & B & \\
    "       & 48447.73816 & 35\,587 &$V$&   15     & 122 & $-$0.5,+0.35  & B & \\
    "       & 48447.81620 & 35\,588 &$V$&   15     & 64  & $-$0.22,+0.14 & B & \\
1991 Jul 10 & 48448.55692 & 35\,600 &$R$&   15     & 99  & $-$0.17,+0.38 & B & 1.5 -- 2.0\\
1992 Jul 27 & 48831.47956 & 41\,834 &$R$&  15,50   & 143 & $-$0.34,+0.5  & C & 1.3 -- 1.8\\
    "       & 48831.55100 & 41\,835 &$R$&  30,50   & 71  & $-$0.5,+0.5   & C & \\
    "       & 48831.61239 & 41\,836 &$R$&  15,50   & 115 & $-$0.5,+0.5   & C & \\
    "       & 48831.67381 & 41\,837 &$R$&  15,50   & 111 & $-$0.5,+0.5   & B & \\
    "       & 48831.73532 & 41\,838 &$R$&  15,50   & 119 & $-$0.5,+0.43  & A & \\
1992 Jul 28 & 48832.55223 & 41\,851 &$R$&  20,50   & 93  & $-$0.2,+0.5   & B & 1.5 -- 2.0\\
    "       & 48832.59537 & 41\,852 &$R$&  15,50   & 124 & $-$0.5,+0.5   & B & \\
    "       & 48832.65663 & 41\,853 &$R$&  15,50   & 163 & $-$0.5,+0.6   & B & \\
1992 Jul 29 & 48833.53336 & 41\,867 &$V$&  20,50   & 67  & $-$0.22,+0.28 & B & 1.3 -- 2.3\\
    "       & 48833.60199 & 41\,868 &$V$&   20     & 68  & $-$0.1,+0.33  & B & \\
    "       & 48833.65432 & 41\,869 &$V$&  20,50   & 103 & $-$0.25,+0.5  & B & \\
    "       & 48833.70111 & 41\,870 &$V$&  20,50   & 114 & $-$0.5,+0.49  & B & \\
    "       & 48833.76214 & 41\,871 &$V$& 20,25,50 & 54  & $-$0.5,+0.18  & B & \\
1992 Jul 30$^{*}$ & 48834.44980 &  $41\,882$  &$R$&   50     &  1  & \multicolumn{1}{c}{$-$0.3} 
& B & 1.3 -- 1.5\\
    "       & 48834.45659 & " &$V$&   25     & 61  & $-$0.2,+0.26  & B & \\
    "       & 48834.52074 & 41\,883 &$V$&  20,50   & 87  & $-$0.15,+0.5  & B & \\
    "       & 48834.56075 & 41\,884 &$V$&  20,50   & 90  & $-$0.5,+0.25  & B & \\
    "       & 48834.64339 & 41\,885 &$V$&  20,50   & 80  & $-$0.16,+0.5  & B & \\
    "       & 48834.68395 & 41\,886 &$V$&  20,50   & 110 & $-$0.5,+0.5   & B & \\
    "       & 48834.74569 & 41\,887 &$V$&   20     & 90  & $-$0.49,+0.39 & B & \\
1998 Jul 26 & 51021.71684 & 77\,488 &$B$&   20     & 203 & $-$0.27,+0.5  & A & 1.2 -- 1.4\\
    "       & 51021.76387 & 77\,489 &$B$&   20     & 295 & $-$0.5,+0.34  & A & \\
1999 Jul 12 & 51372.60399 & 83\,200 &$B$&   20     & 108 & $-$0.25,+0.16 & A & 1.2\\
    "       & 51372.77520 & 83\,202 &$B$&   20     & 164 & $-$0.47,+0.16 & A & \\
1999 Jul 14 & 51374.79735 & 83\,236 &$B$&   20     & 386 & $-$0.55,+0.57 & A & 1.2 -- 1.5\\
2000 Jul 29 & 51755.67684 & 89\,436 &$B$&   20     & 165 & $-$0.29,+0.33 & A & 1.2 -- 1.5\\
2000 Jul 30 & 51756.47832 & 89\,449 &$B$&   20     & 103 & $-$0.25,+0.14 & A & 1.3 -- 1.5\\
    "       & 51756.60539 & 89\,451 &$B$&   20     &  92 & $-$0.18,+0.17 & A & \\
2001 Jun 25 & 52086.72728 & 94\,825 &$B$&   20     & 113 & $-$0.20,+0.23 & A & 1.5 -- 2.0\\
2001 Jun 28 & 52089.67373 & 94\,873 &$B$&   20     & 124 & $-$0.23,+0.24 & B & 1.3 -- 1.8\\
    "       & 52089.80202 & 94\,875 &$B$&   20     &  69 & $-$0.14,+0.22 & B & \\[1ex]
\hline\hline
\multicolumn{8}{l}{{\footnotesize$^{\dagger}$Night Quality: A -- photometric ;
B -- good; C -- poor (large variations and/or clouds).}}\\
\multicolumn{8}{l}{{\footnotesize$^{*}$See text (Section~\ref{sec:dwarfn}).}}\\
\end{tabular}
\end{table*}
\normalsize

\begin{figure}
\centering
\resizebox{\hsize}{!}{\includegraphics{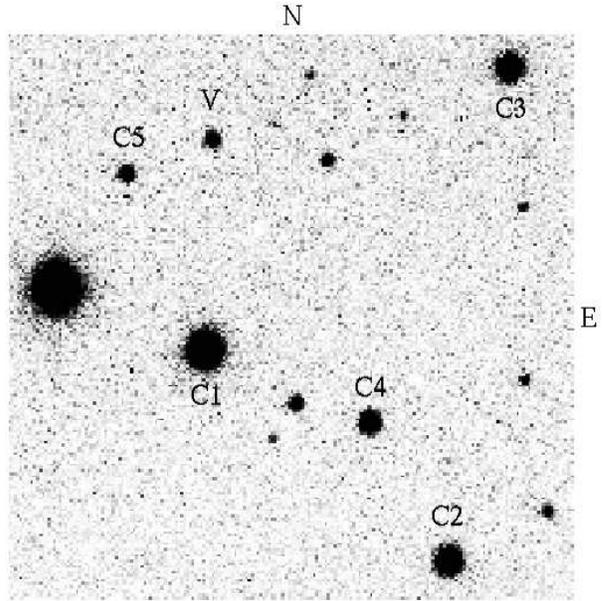}}
\caption{Image of V4140 Sgr (labeled with V) obtained with a 
  CCD camera in the $V$ band in 1992. The reference star (C1) and the
  comparison stars (C2 to C5) used in the differential photometry are
  indicated.  The field orientation in North up and East to the right,
  and its size is $\simeq 2\arcmin \times 2\arcmin$.}
\label{fig:chart}
\end{figure}

The CCD data reduction was done with IRAF\footnote{\emph{Image
    Reduction and Analysis Facility}, a general purpose software
  system for the reduction and analysis of astronomical data. IRAF is
  written and supported by National Optical Astronomy Observatories
  (NOAO) in Tucson, Arizona. NOAO is operated by the Association of
  Universities for Research in Astronomy Inc. (AURA) under cooperative
  agreement with the National Science Foundation.} routines and
included the usual bias and flat-field corrections, and the automatic
identification and removal of cosmic rays. Aperture photometry was
carried out with the IRAF APPHOT package. The fluxes were extracted
for the variable and for 5 field stars using a 6-pixel radius
diaphragm. The sky contribution was estimated from the centroid of the
intensity histogram (or weighted mean) of all pixels within an inner
radius of 7 pixels and a radial width of 10 pixels, centered in each
star. Figure~\ref{fig:chart} shows an image of the field of
V4140 Sgr in which the reference and comparison stars used
for the differential photometry are indicated. Time series were
constructed by computing the magnitude difference of the variable (V)
and other comparison stars (C2 to C5) with respect to the reference
star (C1). The uncertainty in the differential photometry of
V4140 Sgr was estimated from the dispersion in the magnitude
difference for comparison stars of similar brightness and amounts to
0.06, 0.12 and 0.065 mag respectively in the $B$, $V$ and $R$ bands.
The data were transformed from magnitude to a relative flux scale by
assuming a unity flux for the reference star (C1). The reference star
flux was calibrated from observations of standard stars of Graham
(1982) and of blue spectrophotometric standards (Stone \& Baldwin
1983). The photometric calibration program adopts the prescription of
Harris et al. (1981) to solve simultaneously all photometric
coefficients, including first and second-order extinction (e.g.
Jablonski et al.  1994). We used the relations of Lamla (1982) to
transform the $BVR$ magnitudes of the reference star to flux units and
to flux calibrate the light curves of V4140 Sgr.

\section{V4140 Sgr as a dwarf nova}
\label{sec:dwarfn}

\begin{figure}
  \centering \resizebox{\hsize}{!}{\includegraphics{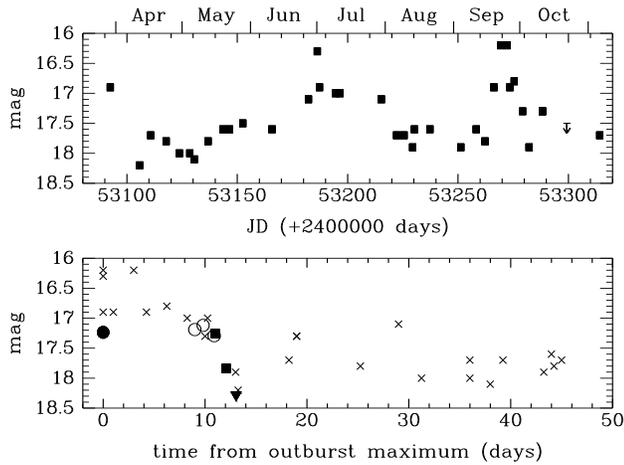}}
\caption{\emph{Top:} The visual historical light curve of V4140 Sgr
  in 2004 constructed from observations made by VSNET amateur
  astronomers. The epoch of the observations are indicated on the top.
  \emph{Bottom:} Superimposed outbursts of V4140 Sgr. Crosses
  indicate visual measurements. Filled symbols are out-of-eclipse mean
  magnitude of the object from the data of 2001 $B$ (circle), 1992 $R$
  (squares) and 1992 $V$ (triangle) (see text).  Open circles are from
  the $V$ measurements of BJS.}
\label{fig:hist}
\end{figure}

The top panel of Fig.~\ref{fig:hist} displays the visual light curve
of V4140 Sgr constructed from observations made by VSNET
amateur astronomers (Kato et al. 2004) during 2004.  The observations
were performed with an unfiltered CCD and reduced to visual magnitude
using the R magnitude sequence\footnote{see http://www.aavso.org
  (AAVSO homepage)}. The star was observed in outburst three times on
a time interval of 7 months.  On the 2004 September outburst
V4140 Sgr showed superhumps with a period 1.6 per cent longer
than the orbital period (Uemura 2004; Uemura et al. 2004, in
preparation), thereby confirming that V4140 Sgr is not only a
dwarf nova, but also an SU UMa star. The outburst of 2004
June was possibly another superoutburst.  The maximum of the 2004
April outburst is fainter than those of the following superoutbursts
by $\simeq 0.7$ mag -- in agreement with the expected for a regular
outburst.  From Fig.2 we infer a mean time interval between outbursts
of $~ 80-90$ days and an outburst duration of $5-10$ days, typical of
dwarf novae (Warner 1995).

The outbursts of V4140 Sgr have remarkably low amplitude in
comparison with bona fide dwarf novae. The maximum of the 2004 Sep
superoutburst is only about 2 mags brighter than the typical quiescent
level of $V \simeq 18.0$ mag. Accordingly, the regular outburst (2004
Apr) has an amplitude of only $\simeq 1$ mag. These results suggest
that the quiescent disc of V4140 Sgr may be significantly
brighter than those of the prototype short period dwarf novae
OY Car and Z Cha.

\begin{figure}
\centering
\resizebox{\hsize}{!}{\includegraphics{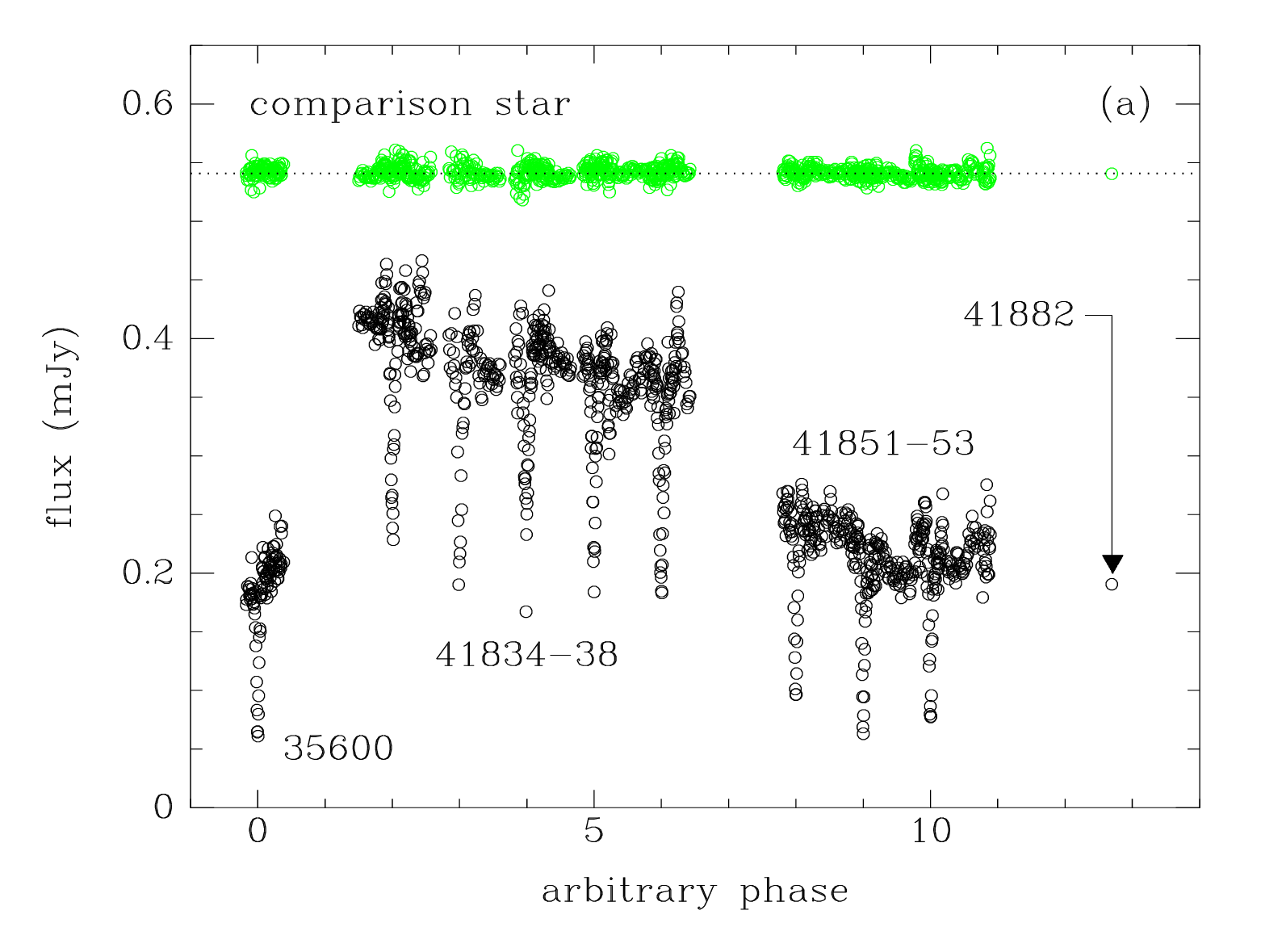}}
\resizebox{\hsize}{!}{\includegraphics{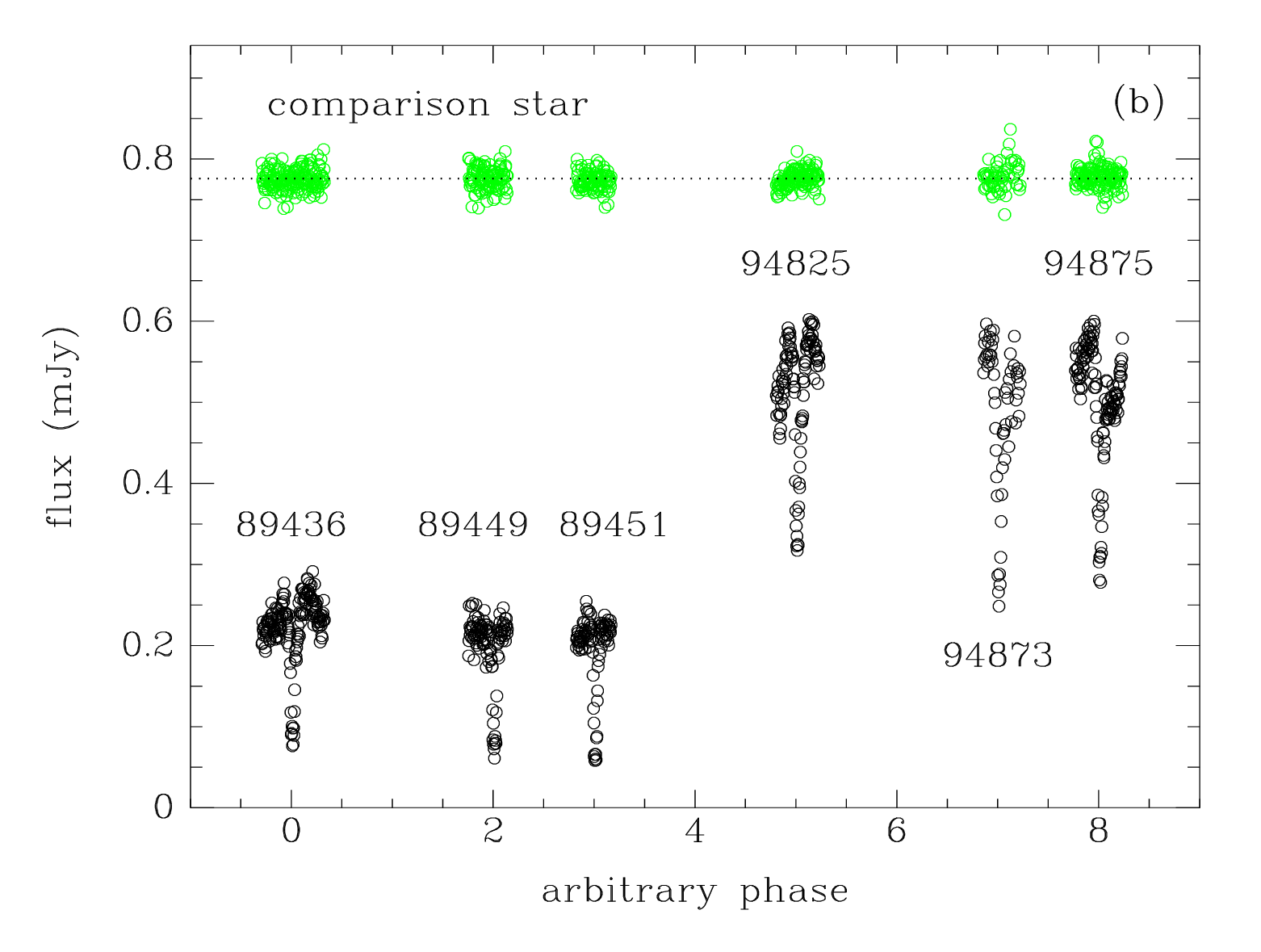}}
\caption{(a) $R$ band observations of V4140 Sgr in the decline 
  from an outburst in 1992 July. The labels indicate the cycles of the
  observations (see Table~\ref{tab:journal} for details). The only
  point in 1992 July 30th refer to a single $R$ band image at that
  night. The gray points show the light curve of a field comparison
  star. (b) $B$ band observations of V4140 Sgr in outburst in
  2001 June in comparison to the quiescence data of 2000 July. The
  separation of the light curves in the $x$-axis is arbitrary in both
  panels and was chosen for visualization purposes only.}
\label{fig:obs_out}
\end{figure}

The top panel of Fig.~\ref{fig:obs_out} shows our $R$ band
observations of V4140 Sgr in 1992. The brightness of the star
reduced by a factor 2 on a time scale of one night, indicating that
the object was caught in the decline from an outburst. The eclipse
runs are arbitrarily shown in consecutive binary cycles for a better
visualization. Therefore, the real time span between eclipses on
different nights is larger than indicated in the figure, except for
the truly consecutive cycles. The eclipse in the night of 1991 July
10th (cycle $35\,600$) is representative of the quiescent state. The
data point in 1992 July 30th (cycle 41882) is from a single exposure
in the $R$ band taken outside of eclipse (orbital phase $-0.3$) and is
consistent with the trend in brightness decline inferred from the
observations of the previous nights. The object was also caught in
outburst in 2001, from observations in the $B$ band. The lower panel
of Fig.~\ref{fig:obs_out} shows the brightness increase of
V4140 Sgr in 2001 June in comparison to observations done in
a previous year, when the object was in quiescence. The object shows
brightness variations by a factor of $\simeq 2.5$ in the two observed
outbursts, confirming that the normal outbursts have small amplitude
($\Delta$mag = 1 mag). In both cases, the outburst eclipses are
relatively shallow, with a depth comparable to that in quiescence,
indicating that the brightness increase is mainly from the outer and
only partially eclipsed disc regions.

The lower panel of Fig.~\ref{fig:hist} presents the superimposed
outbursts of V4140 Sgr. We use the decline branch of the
outburst to align the incomplete outbursts in time and we set the zero
time as that corresponding to the maximum brightness of those outburst
with full observing coverage. Crosses represent VSNET observations
(upper panel of Fig.~\ref{fig:hist}) in visual magnitude.  Filled
symbols show measurements of the out-of-eclipse mean magnitude of our
data: $B$ in outburst in 2001 (circle), $R$ (square) and $V$
(triangle) in decline from the 1992 outburst (see
Table~\ref{tab:journal} for details). The $V$ and $R$ points were
spaced using the mean interval between the observations, and were
displaced in time to fit the decline feature of the VSNET outbursts
for comparison. Because the outburst stage of our $B$ measurement are
unknown, we plotted this point at zero time for comparison purposes.
The open circles refer to $V$ out-of-eclipse magnitudes measurements
of BJS in 1988, when the object was probably also in outburst (see the
discussion of Baptista \& Steiner 1991). These points were also
aligned to the early decline region of the VSNET outbursts.

The slope of the decline branch of our data in $V$ and $R$ bands is in
accordance with that derived from the VSNET data. Our $B$ band
measurement is probably representative of the flat-top branch of
maximum light of a regular outburst, since the star kept that same
brightness level for at least 3 nights. All of our data shown in
Fig.~\ref{fig:obs_out} are consistent with some stage of the outburst
or decline from an outburst of V4140 Sgr in comparison to the
observed time scales and amplitudes of VSNET observations. The BJS
points could also correspond to early decline phase from outburst
maximum.

The outburst decay time scale seen in the lower panel of
Fig.~\ref{fig:hist} is of order of $1-2$ days, much smaller than the
time scale of hundreds of days of the transition to the low states of
VY Scl-type stars or tens of days seen in fading episodes of some
dwarf novae (Kato 2004). In order to determine the decay time scale
$\tau_{{\rm d}}$ (in days\,mag$^{-1}$), we fit an exponential function
to the $R$ band data of 1992, excluding the times in which the object
was in eclipse. The inferred values of $\tau_{{\rm d}} \simeq 1.2$
d\,mag$^{-1}$ is in agreement with those of SU UMa dwarf
novae in a $\log \tau_{{\rm d}} \times \log P_{{\rm orb}}$ diagram
(Warner 1995).

The observed amplitude of the superoutbursts ($\Delta$mag $\simeq 2$)
and the expected amplitude of the regular outbursts ($\Delta$mag
$\simeq 1$) of V4140 Sgr are in clear disagreement with the
typical values for dwarf novae stars. If we consider an outburst
amplitude in the range $A_{n} = 1 - 2$ mag and a recurrence time
interval in the range $T_{n} = 80 - 90$ days, the object falls far
away from the expected Kukarkin-Parenago relation in an amplitude
versus recurrence time for dwarf novae outbursts (Warner 1995).

\section{Data analysis}
\label{sec:data_an}

\subsection{The revised orbital parameters}
\label{sub:orbit}

We found an error in the determination of the binary parameters of
BJS, caused by the use of a wrong relation for the width of the white
dwarf eclipse (see Fig. 6 of BJS). Here the binary parameters of V4140
Sgr are revised. We adopted the same observed parameters of BJS: a
width of the white dwarf eclipse of $\Delta \phi = 0.0378 \pm 0.0005$
cycles and a mean duration of the white dwarf ingress/egress feature
of $\Delta_{{\rm wd}} = 0.009 \pm 0.001$ cycles. The value of $\Delta
\phi$ derived from our data is the same as above within the
uncertainties.

The basis of the photometric binary parameters determination method
consists in the measurement of white dwarf contact phases. Detailed
descriptions of the method can be found in BJS and Baptista \&
Catal\'{a}n (2000). A Monte Carlo propagation code was used to
estimate the errors in the calculated parameters. The values of the
input parameters are independently varied according to Gaussian
distributions with standard deviations equal to the uncertainty of
those parameters. The new results are listed in Table~\ref{tab:param}
with their 1-$\sigma$ errors. The parameters of BJS are also listed
for comparison. $K_{1}$ and $K_{2}$ are the expected values of the
primary and secondary star radial velocity semi-amplitude,
respectively, as derived from our photometric model.

\begin{table}
\centering
\caption{Comparison of the orbital parameters} \label{tab:param}
\begin{tabular}{@{}lll@{}}
\hline\hline
                  & Present work      & BJS (1989)          \\[1ex] \hline
$q$               & $0.125\pm0.015$   & $0.15\pm0.03$       \\[1ex]
$i$               & $80.2^{\circ}\pm0.5^{\circ}$ & $79.4^{\circ}\pm0.9^{\circ}$ \\[1ex]
$M_{1}/M_{\odot}$ & $0.73\pm0.08$     & $0.44\pm0.07$       \\[1ex]
$M_{2}/M_{\odot}$ & $0.092\pm0.016$   & $0.07\pm0.02$       \\[1ex]
$R_{1}/R_{\odot}$ & $0.0108\pm0.0008$ & $0.014\pm0.002$     \\[1ex]
$R_{2}/R_{\odot}$ & $0.136\pm0.008$   & $0.12\pm0.01$       \\[1ex]
$a/R_{\odot}$     & $0.61\pm0.05$     & $0.52\pm0.03$       \\[1ex]
$K_{1}$ (km/s)    & $55\pm7$          &\ \ \ \ \ \  --      \\[1ex]
$K_{2}$ (km/s)    & $442\pm15$        & $370\pm20$          \\[1ex] \hline\hline
\end{tabular}
\end{table}

\begin{figure}
\centering
\resizebox{\hsize}{!}{\includegraphics{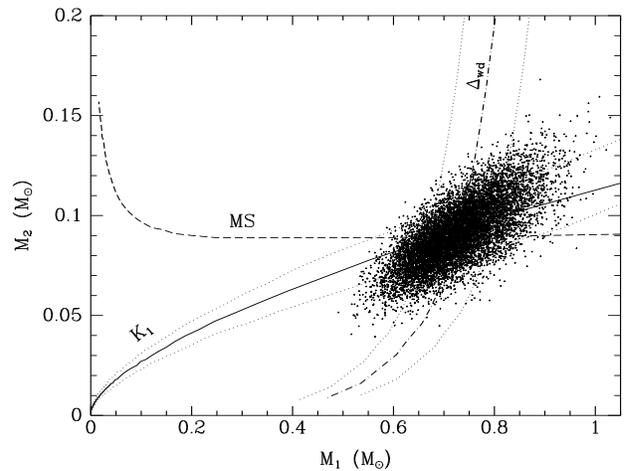}}
\caption{The primary-secondary mass diagram for V4140 Sgr. 
  The cloud of points is the result of $10^{4}$ Monte Carlo
  simulations with the data points. The dashed line labeled as MS
  depicts the empirical mass-radius relation of Caillaut-Patterson for
  main sequence stars.  The solid line is the relation derived from
  the primary radial velocity semi-amplitude $K_{1}$ obtained by
  Mukai, Corbet \& Smale (1988) and the dot-dashed line is the
  $M_{1}(q)$ relation obtained by the measurement of $\Delta_{{\rm
      wd}}$. The dotted lines correspond to the uncertainties at the
  1-$\sigma$.}
\label{fig:mass_diag}
\end{figure}

The primary-secondary mass diagram for V4140 Sgr can be seen
in Figure~\ref{fig:mass_diag}. The cloud of points was obtained from a
set of $10^{4}$ trials using the Monte Carlo code.  The highest
concentration of points indicates the region of most probable
solutions.  BJS found $R_{2} = 0.12 R_{\odot}$, suggesting that the
secondary star was oversized with respect to main sequence stars of
same mass. Our revised determination leads to $M_{2} = 0.09 M_{\odot}$
and $R_{2} = 0.136 R_{\odot}$, consistent with the empirical
mass-radius relation of Caillaut \& Patterson (1990) at the 1-$\sigma$
level. The revised secondary mass is also consistent with the results
of Smith \& Dhillon (1998), who found that the secondary stars of CV
are indistinguishable from isolated main sequence stars of same mass.

The new value of the mass ratio $q (= M_{2}/M_{1})$ differs from the
previous value by less than 20\%, while the difference in the
inclination $i$ is less than 1\%. The largest change occurs in the
primary mass. The previous value of $M_{1} = 0.44 M_{\odot}$ was much
lower than the observed mean white dwarf mass in CVs below the gap
($\overline{M_{1}} = 0.69 \pm 0.13 M_{\odot}$, Smith \& Dhillon 1998;
$\overline{M_{1}} = 0.66 \pm 0.01 M_{\odot}$, Webbink 1990). The
revised value of $M_{1} = 0.73 M_{\odot}$ is consistent with the mean
white dwarf of Smith \& Dhillon (1998). The calculated values of the
orbital separation $a$ and the predicted $K_{2}$ increased by 18\%. A
detailed discussion about the uncertainties affecting the photometric
model developed here can be found in BJS.

\subsection{The eclipse maps}
\label{sub:emm}

The eclipse mapping method (EMM) is an inverse problem resolution
technique that uses the information contained in the shape of the
eclipse light curve to reconstruct the surface brightness distribution
of the accretion disc through an entropy maximization procedure
(Skilling \& Bryan 1984). For details of the method, the reader is
referred to Horne (1985), Rutten et al. (1992a), Baptista \& Steiner
(1993) and Baptista (2001).

The individual light curves were phase folded according to linear plus
sinusoidal ephemeris of Baptista et al. (2003)

\begin{eqnarray}
T_{{\rm mid}} & = & {\rm BJDD}\ 2\,446\,261.671\,50 + 0.061\,429\,6757\,E + 
\nonumber \\
 & & + 19 \times 10^{-5}\,\cos \left[ 2\pi \left( \frac{E - 3 \times 10^{3}}
{41 \times 10^{3}} \right) \right]\ \ \ \ {\rm d.}
\label{eq:ephem}
\end{eqnarray}

In order to improve the signal-to-noise ratio and to minimize the
influence of the characteristic flickering of CVs, the individual
light curves were grouped per band and brightness state to produce
average eclipse light curves. For each light curve, the data were
divided in phase bins of 0.003 cycles and the median was computed for
each bin. The median of the absolute deviations with respect to the
median was taken as the corresponding uncertainty on each bin.
Table~\ref{tab:id} summarizes the data used for the EMM, indicating
the dates, passbands and brightness state of each average light curve
(the details can be seen in Table~\ref{tab:journal}). The last column
lists the label used for further identification of the average light
curves.

\begin{table}
\centering
\caption{Identification of the average light curves used in EMM} \label{tab:id}
\begin{tabular}{@{}cccl@{}}
\hline\hline
Date                   & Passband & Brightness State & Identification\\[1ex] \hline
1998, 1999, 2000$^{*}$ &   $B$    & quiescence       & \emph{Bquies} \\
2001$^{*}$             &   $B$    & outburst         & \emph{Bout}   \\
1991$^{*}$             &   $V$    & quiescence       & \emph{Vquies} \\
1992$^{*}$             &   $V$    & outburst decline & \emph{Vdecl}  \\
1991$^{*}$             &   $R$    & quiescence       & \emph{Rquies} \\
1992 Jul 27            &   $R$    & outburst         & \emph{Rout}   \\
1992 Jul 28            &   $R$    & outburst decline & \emph{Rdecl}  \\[1ex] \hline\hline
\multicolumn{4}{l}{{\footnotesize$^{*}$Corresponds to all cycles of the year(s) in
that passband}}\\
\multicolumn{4}{l}{{\footnotesize (see Table~\ref{tab:journal} for details).}}\\
\end{tabular}
\end{table}

Out-of-eclipse brightness variations are not accounted for the EMM,
which assumes that all brightness variations are due to occultation of
parts of the disc by the secondary star. The out-of-eclipse brightness
variations were removed from the average light curves by fitting a
spline function to the phases outside eclipse. The light curve was
divided by the fitted spline and the result was scaled to the value of
the spline at phase zero. This procedure removes the orbital
variations with a minor effect on the eclipse shape itself. To speed
the convergence of the EMM, all points outside of phase range ($-
0.15, +0.15$) were removed. This does not affect the analysis since
the points ruled out are far from the narrow eclipse of the accretion
disc of V4140 Sgr.

The eclipse geometry was defined by the mass ratio $q = 0.125$ and
inclination $i = 80.2^{\circ}$ (Section~\ref{sub:orbit}), which
corresponds to a value of $\Delta \phi = 0.0378$ cycle for the width
of the white dwarf eclipse. As mentioned above, the average light
curves were phase folded by Eq.~\ref{eq:ephem}, therefore the phase of
inferior conjunction is $\phi_{0} = 0$. This combination of parameters
ensures that the white dwarf is at the centre of the map. A flat grid
of $51 \times 51$ pixels centered in the white dwarf with side
$2R_{L_{1}}$ was adopted as eclipse map. For V4140 Sgr, a
value of $R_{L_{1}} = 0.424 R_{\odot}$ is obtained.

\begin{figure}
\centering
\resizebox{\hsize}{!}{\includegraphics{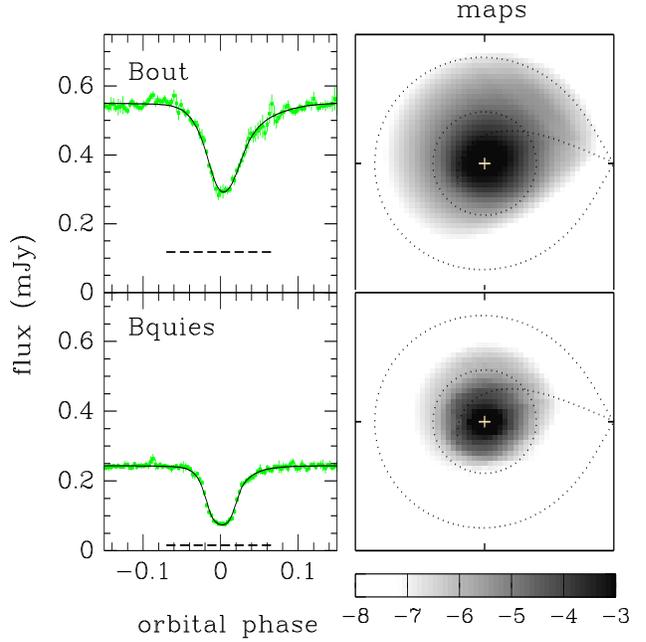}}
\caption{\emph{Left:} Average light curves of V4140 Sgr in 
  the $B$ band with respective error bars (dots) and the model light
  curves (solid lines). A dashed line indicates the uneclipsed
  component in each case. \emph{Right:} Surface brightness
  distributions of the accretion disc of V4140 Sgr in a
  logarithmic grey-scale.  Brighter regions are indicated in black and
  the fainter regions in white. A cross marks the centre of the disc.
  The dotted lines show the Roche lobe, the ballistic stream
  trajectory (Flannery 1975) and a disc radius of $R_{{\rm d}} = 0.4
  R_{L_{1}}$ (see text for explanation).  The grey-scale bar
  corresponds to a linear scale in log of intensity from $-8$ to
  $-3$.}
\label{fig:map1}
\end{figure}

\begin{figure}
\centering
\resizebox{\hsize}{!}{\includegraphics{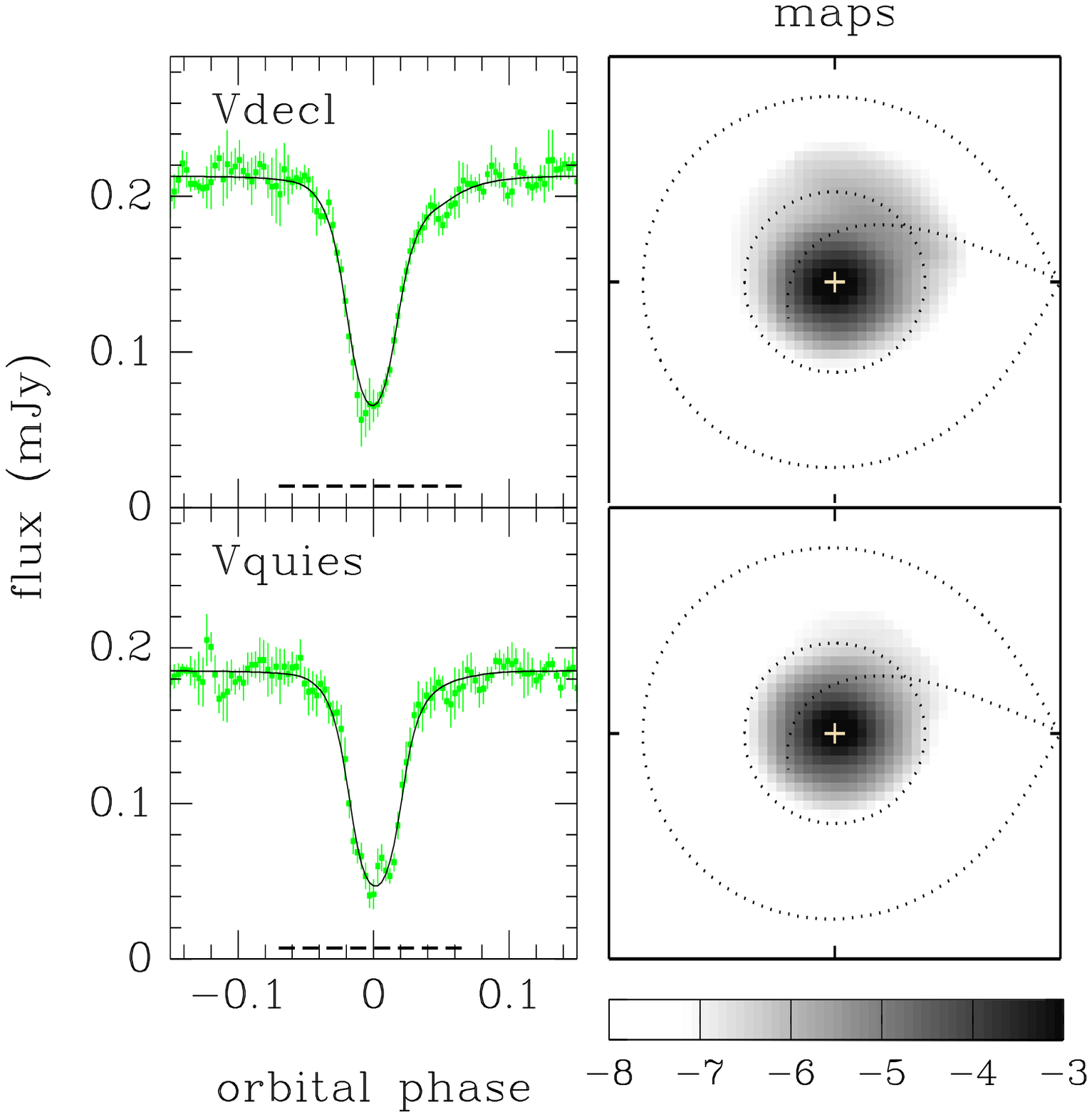}}
\caption{\emph{Left:} Average light curves of V4140 Sgr in 
  the $V$ band with respective error bars (dots) and the model light
  curves (solid lines). \emph{Right:} Surface brightness distributions
  of the accretion disc of V4140 Sgr in a logarithmic
  grey-scale.  The notation is the same as in Fig.~\ref{fig:map2}.}
\label{fig:map2}
\end{figure}

\begin{figure}
\centering
\resizebox{\hsize}{!}{\includegraphics{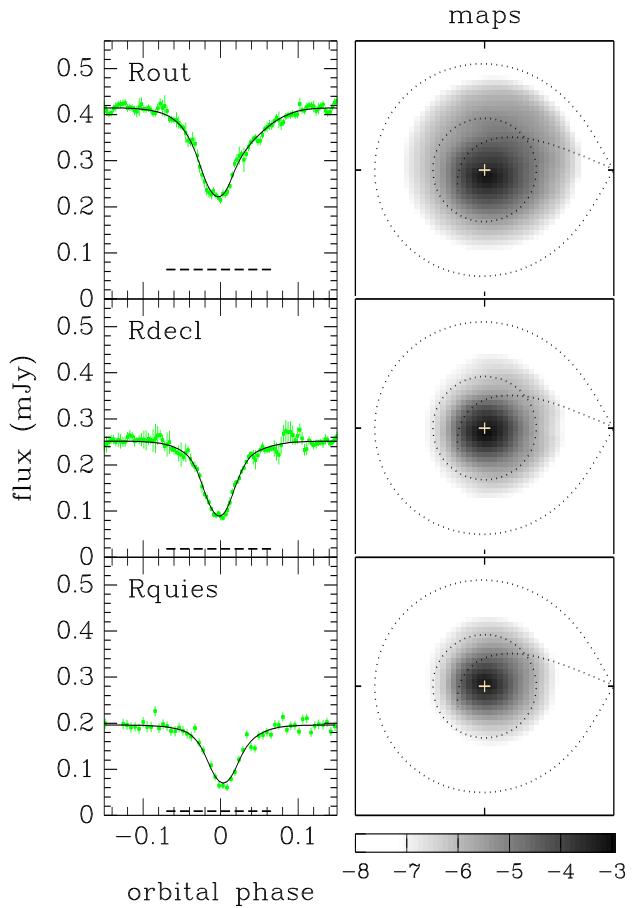}} 
\caption{\emph{Left:} Average light curves of V4140 Sgr in 
  the $R$ band with respective error bars (dots) and the model light
  curves (solid lines).  \emph{Right:} Surface brightness
  distributions of the accretion disc of V4140 Sgr in a
  logarithmic grey-scale. The notation is the same as in
  Fig.~\ref{fig:map2}.}
\label{fig:map3}
\end{figure}

The average light curves were then analyzed with EMM to reconstruct
the surface brightness distribution of the accretion disc of
V4140 Sgr and to obtain the additional flux of the uneclipsed
component. The latter accounts for all light that is not contained in
the eclipse map and can be obtained from the entropy function (Rutten
et al. 1992a). A default map of limited azimuthal smearing (Rutten et
al. 1992a) was adopted, which is more appropriate for recovering
asymmetric structures than the original default of full azimuthal
smearing (Baptista et al. 1996). The map are shown in
Figs.~\ref{fig:map1}, \ref{fig:map2} and \ref{fig:map3}. The
identification of the average light curves follows the nomenclature of
Table~\ref{tab:id}. The quiescent disc radius was estimated from the
position of the bright spot in the $V$ quiescence state map (the
centroid of the asymmetry in the \emph{Vquies} eclipse map) to be
$R_{{\rm d}} \simeq 0.4 R_{L_{1}}$.

Table~\ref{tab:unecl} summarizes the fractional contribution of the
uneclipsed component by passband and by brightness state. The
contribution of the uneclipsed component increases significantly in
outburst, reaching more than 20\% of the total flux in the $B$ band.
The two natural hypotheses to explain the observed increase of the
uneclipsed flux during the outburst are the appearance of a wind from
the disc and the presence of a flared accretion disc (Baptista \&
Catal\'{a}n 2001).

\begin{table}
\centering
\caption{Fractional contribution of the uneclipsed component} \label{tab:unecl}
\begin{tabular}{@{}cccc@{}}
\hline\hline
Passband  &       \multicolumn{3}{c}{Brightness state}           \\
          & \emph{outburst} & \emph{decline} & \emph{quiescence} \\[1ex] \hline           
 $B$      &      0.21       &       --       &       0.07        \\
 $V$      &       --        &      0.07      &       0.04        \\
 $R$      &      0.15       &      0.07      &       0.05        \\[1ex] \hline       
\end{tabular}
\end{table}

The average quiescent light curves show narrow, smooth and slightly
asymmetric eclipses, which map into smooth brightness distributions
with maximum light at disc centre that are slightly brighter at the
position of the bright spot at disc rim. During outburst, the eclipses
become shallower and more asymmetric towards the egress, resulting in
eclipse maps with increased brightness in the outer parts of the
accretion disc and larger asymmetries at the bright spot position.
There is no evidence of a front-to-back brightness contrast in the
eclipse maps, which argues against a significant flaring of the
accretion disc at this stage\footnote{If the disc is flared, the disc
  side farther away from the secondary star (the ``back'' side) is
  made apparently brighter than the disc side closest to the secondary
  (the ``front'' side) because it is seen at a lower inclination angle
  and has larger effective area.}. The $V$ band light curve of 1992
(\emph{Vdecl}, see Table~\ref{tab:id}) represents a brightness state
just after an outburst and may not correspond to the typical quiescent
level. Thus, for the following analysis we adopted the light curve and
eclipse map of 1991 as representative of the quiescent state in the
$V$ band (and labeled as \emph{Vquies} in consequence).

The uncertainty of each pixel in the eclipse maps was estimated by a
bootstrap procedure (Efron 1982). We generated 20 light curves for
each average curve identified in Table~\ref{tab:id} and applied the
EMM to these curves to obtain eclipse maps. The uncertainty of each
pixel was taken as the absolute deviations of the sample of 20
bootstrap maps with respect to the intensity of the real map.

\section{Results}
\label{sec:resul}

\subsection{A distance estimate}
\label{sub:distan}

The quiescent eclipse maps in the $V$ and $R$ bands allow us to
estimate the distance if the disc is optically thick (Bruch et al.
1996). This determination is quite similar to the distance estimate of
an open cluster, which is done by fitting the standard main sequence
relation to the main sequence stars of the cluster in a color-color
diagram. However, each map pixel represented in this diagram has the
same solid angle, while real stars not.  Therefore, for a correct
application of the method, a calibration of the magnitude in terms of
surface brightness is required.

It is difficult to investigate the emission properties of an accretion
disc with a $(B-V) \times (V-R)$ color-color diagram because of the
proximity of the optically thick (blackbody and main sequence) and
optically thin relations in this diagram. Previous similar studies
(Bruch et al. 1996; Baptista et al. 1996) were based on a $(U-B)
\times (B-V)$ diagram, in which the optically thin and optically thick
emission relations are clearly separated.

The Barnes-Evans empirical relation (Barnes \& Evans 1976; Barnes et
al. 1978) relates the surface brightness parameter $F_{V}$ to the
color index $(V-R)$, the apparent visual magnitude and the angular
diameter. The EMM program yields the intensity of each pixel as seen
under the standard condition (a distance of 1 kpc, a Roche lobe radius
of $1 R_{\odot}$, and a face-on disc, i.e., $\cos i = 1$).  Because
the $(V-R)$ color index is known for each pixel, the Barnes-Evans
relation allows to evaluate its apparent magnitude for an angular
diameter calculated under standard conditions. The difference between
this value and the value derived directly from the intensities allows
us to estimate the distance. Note again that the method assumes an
optically thick disc emission. If the disc is optically thin, the
inferred distance will be wrong.

\begin{figure}
\centering
\resizebox{\hsize}{!}{\includegraphics{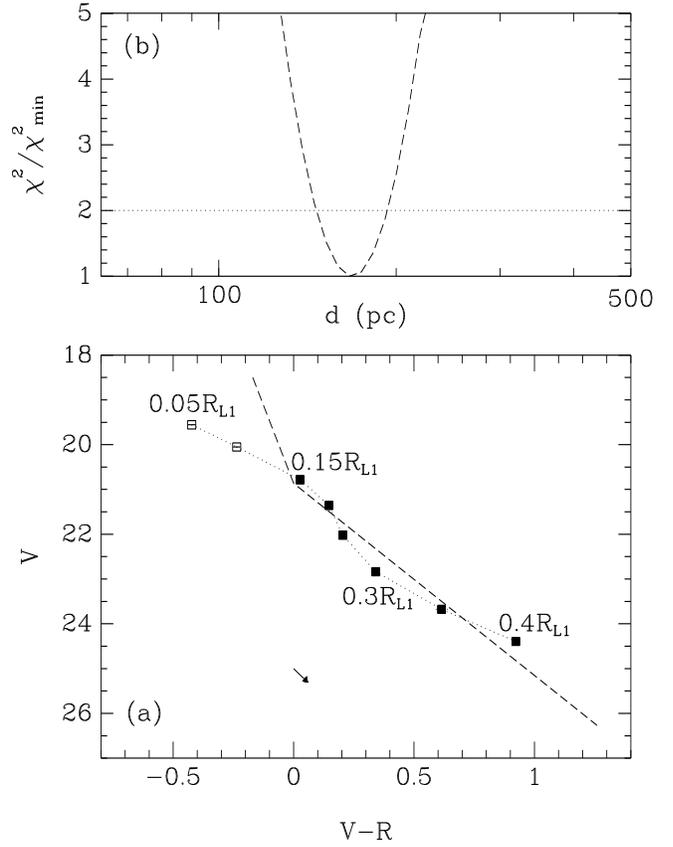}} 
\caption{(a) The color-magnitude diagram used to calculate the distance 
  to V4140 Sgr. Filled squares show the average color and
  magnitude of pixels in radial bins of width $0.05 R_{L_{1}}$ of the
  $V$ and $R$ eclipse maps in quiescence (\emph{Vquies} and
  \emph{Rquies} maps), starting from disc centre ($0.05 R_{L_{1}}$) to
  the outer regions of the disc ($0.4 R_{L_{1}}$). The dashed line is
  the best fit to the Barnes-Evans relation (Barnes, Evans \& Moffet
  1978), and corresponds to a vertical displacement of $\Delta V =
  0.1$ mag of the data points. The first two points (open squares)
  were discarded from the fit because of the central intensity
  lowering effect of the maximum entropy method. A reddening vector
  corresponding to $d = 170 \pm 30$~pc ($E(B-V) = 0.07$) is shown for
  reference. (b) A $\chi^{2}$ versus distance (or versus vertical
  displacement $\Delta V$) diagram is used to find the best fit to the
  Barnes-Evans relation. The dotted line indicates the $\Delta
  \chi^{2}$ level used to estimate the error in $d$.}
\label{fig:dist}
\end{figure}

The lower panel of Fig.~\ref{fig:dist} shows the color-magnitude
diagram used to calculate the distance to V4140 Sgr. The
uncertainty of the points (filled and open squares) are obtained from
the uncertainty in the intensities of individual pixels (see
Section~\ref{sub:emm}) and are typically smaller than the plotted
symbols. A reddening of $E(B-V)=0.4$ mag\,kpc$^{-1}$ is estimated for
V4140 Sgr from the galactic interstellar extinction contour
maps of Lucke (1978). The first two points were discarded from the fit
because of the central intensity lowering effect of the maximum
entropy method (e.g. Bruch et al. 1996). Assuming an optically thick
emission for the disc, a distance of $d = 170 \pm 30$~pc is obtained.
The top panel of Fig.~\ref{fig:dist} presents the $\chi^{2}$ versus
distance diagram used to find the best-fit to the Barnes-Evans
relation.

\subsection{Radial temperature distributions}
\label{sub:distrib}

A basic prediction of the accretion disc theory is that the effective
temperature of a steady-state optically thick disc depends on the
distance from the disc centre $r$ as

\begin{equation}
T^{4}_{{\rm eff}} = \frac{3 G M_{1} \dot{M}}{8 \pi \sigma R^{3}_{1}}
\left( \frac{r}{R_{1}} \right)^{-3} \left[ 1 - \left( \frac{R_{1}}{r}
\right)^{\frac{1}{2}} \right] 
\label{eq:teff} 
\end{equation}

\noindent
where $\dot{M}$ is the mass transfer rate and the remaining symbols
have their usual meaning. This expression is derived from the energy
dissipation rate in the disc by unit surface area (Frank et al. 1992).

\begin{figure}
\centering
\resizebox{\hsize}{!}{\includegraphics{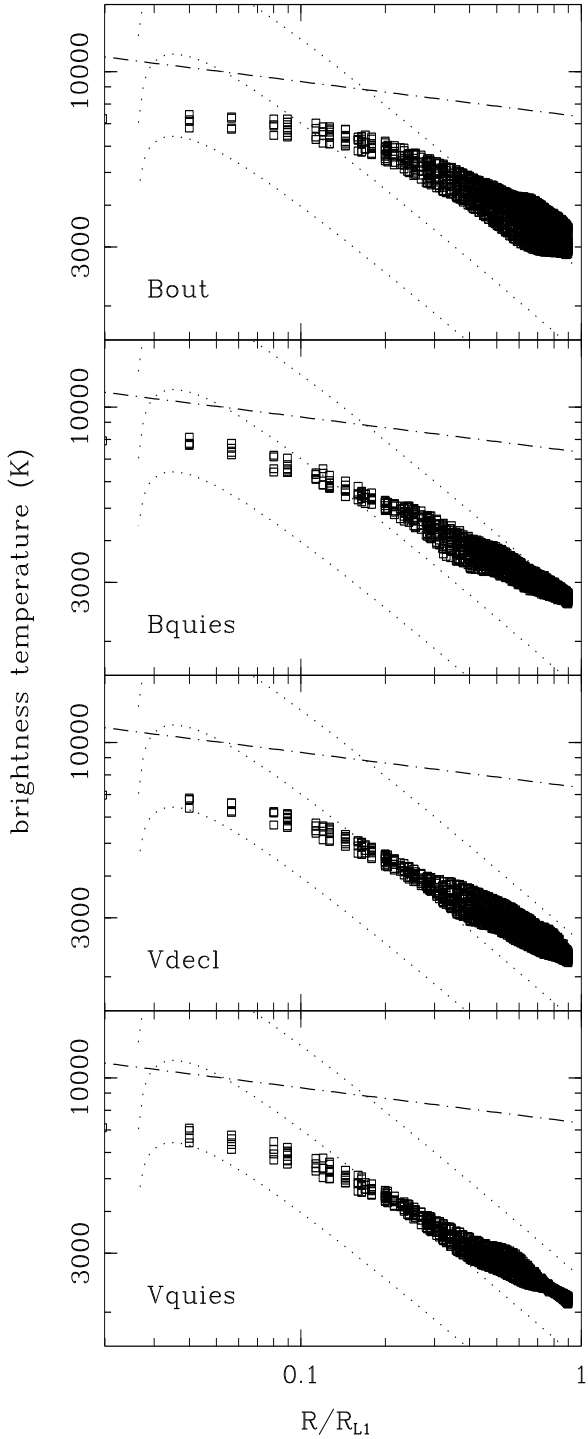}} 
\caption{Radial brightness temperature distributions for the $B$ 
  and $V$ disc maps in outburst, decline and quiescence states. The
  individual pixels of the disc maps are shown as open squares. Dotted
  lines correspond to steady-state optically thick disc models for
  mass accretion rates of $10^{-10}$, $10^{-11}$ and $10^{-12}
  M_{\odot}$yr$^{-1}$, from top to bottom respectively. A dot-dashed
  line shows the critical effective temperature $T_{{\rm crit}}$ below
  which the disc gas should remain to allow the thermal instability,
  and above which the gas should stay while in outburst (e.g. Warner
  1995).}
\label{fig:distr1}
\end{figure}

\begin{figure}
\centering
\resizebox{\hsize}{!}{\includegraphics{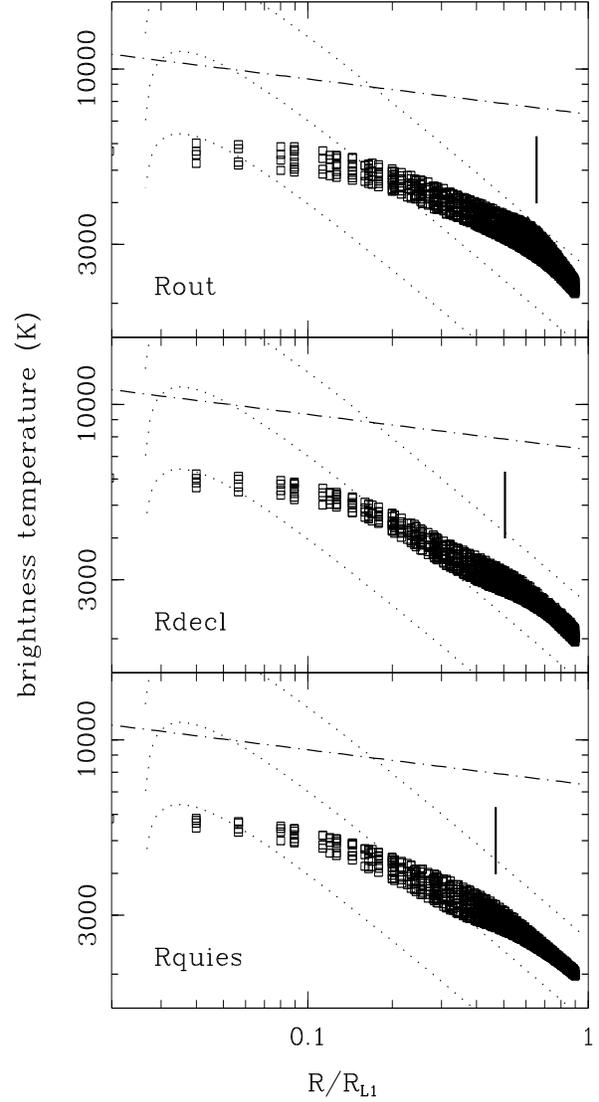}}
\caption{Radial brightness temperature distributions for the $R$ disc 
  maps in outburst, decline and quiescence states. The notation is the
  same as in Fig.~\ref{fig:distr1}. The vertical ticks in the panel
  are used to estimate a cooling wave velocity of $\sim 0.5$ km/s.}
\label{fig:distr2}
\end{figure}

The intensity maps were converted to blackbody brightness temperature
to facilitate its comparison with accretion disc theoretical models.
The radial brightness temperature distributions are presented in a
logarithmic scale in Figs.~\ref{fig:distr1} and \ref{fig:distr2}. The
solid angle used to evaluate the brightness temperature corresponds to
a distance of 170 pc. For each distribution, steady-state optically
thick disc models for different values of $\dot{M}$
(Eq.~\ref{eq:teff}) and the critical effective temperature of the DI
model (Eq.~\ref{eq:tcrit}) are plotted for comparison.

It is important to point out here that the brightness temperature is a
physical quantity different from the effective temperature given by
Eqs.~\ref{eq:teff} and \ref{eq:tcrit}. The relation between those two
quantities is non-trivial, as discussed by Baptista et al. (1996), and
can only be properly obtained by constructing self-consistent models
of the vertical structure of the disc. This is beyond the scope of
this work. On the other hand, the comparison with Eqs.~\ref{eq:teff}
and \ref{eq:tcrit} can be satisfactorily applied to the optically
thick disc regions, where the brightness temperature is expected to be
close to the effective temperature. Therefore, our analysis is only a
good indicator of the disc behavior if the optically thick disc
emission hypothesis is valid.

The temperatures range from $3\,000$ K in the outer disc regions ($r
\sim 0.4 R_{L_{1}}$) to $\sim 6\,000$ K in the inner disc regions ($r
\sim 0.1 R_{L_{1}}$). The temperature distributions, both in
quiescence and in outburst, are flatter than the $T \propto
r^{-\frac{3}{4}}$ law expected for optically thick steady-state
accretion discs. The derived disc temperatures are systematically
lower than the critical temperature $T_{{\rm crit}}$ below which the
disc gas should remain to allow outbursts (according to the DI model),
both in quiescence and in outburst. The outburst occurs mainly with a
significant increase in brightness of the intermediate and outer disc
regions, the temperature of which remain below $T_{{\rm crit}}$ -- in
contrast with the expectation of the DI model.

We estimate the velocity of the inward traveling cooling wave along
the decline by tracing the radial position of a reference, arbitrary
intensity level in the radial intensity distribution. The reference
intensity level corresponds to a brightness temperature of $2\,950$ K.
Vertical ticks mark the radial position of the reference intensity
level along decline in Fig.~\ref{fig:distr2}. Combining the change in
position $\Delta r$ with the time interval between these eclipse maps
(\emph{Rout} and \emph{Rdecl} maps), we infer a cooling wave velocity
of $\simeq 0.5$ km/s, of the same order of the velocity estimated for
the cooling wave in EX Dra (Baptista \& Catal\'{a}n 2001).

Assuming a steady-state disc model, the inferred mass accretion rate
is $10^{-11.6 \pm 0.3} M_{\odot}$yr$^{-1}$ at $r = 0.1 R_{L_{1}}$ and
$10^{-10.8 \pm 0.2} M_{\odot}$yr$^{-1}$ at $r = 0.4 R_{L_{1}}$ for the
quiescent state. At the decline from outburst maximum, the rate is
$10^{-11.6 \pm 0.2} M_{\odot}$yr$^{-1}$ at $r = 0.1 R_{L_{1}}$ and
$10^{-10.9 \pm 0.2} M_{\odot}$yr$^{-1}$ at $r = 0.4 R_{L_{1}}$. For
these two brightness states, the calculated rates are the same at all
passbands under the uncertainties. For the outburst state, the
inferred values of $\dot{M}$ are systematically different for the $B$
and $R$ bands, probaly reflecting the fact that they correspond to
different outbursts. The mass accretion rates inferred from
\emph{Bout} are $10^{-11.4 \pm 0.2} M_{\odot}$yr$^{-1}$ ($r = 0.1
R_{L_{1}}$) and $10^{-10.4 \pm 0.2} M_{\odot}$yr$^{-1}$ ($r = 0.4
R_{L_{1}}$), where for \emph{Rout} the numbers are $10^{-11.7 \pm 0.2}
M_{\odot}$yr$^{-1}$ ($r = 0.1 R_{L_{1}}$) and $10^{-10.7 \pm 0.2}
M_{\odot}$yr$^{-1}$ ($r = 0.4 R_{L_{1}}$). Since the accretion disc is
seemingly not in a stationary regime during outburst, these mass
accretion rates are only illustrative and should be looked at with
some reservation.

\section{Discussion}
\label{sec:disc}

V4140 Sgr shows a peculiar behavior when its quiescent and outburst
radial temperature distributions are compared with those of other
dwarf novae of similar orbital period. In order to quantify the
differences, we fitted a relation of the type $T_{{\rm b}}(r) \propto
r^{-\beta}$ to the radial temperature distributions in the range
$0.1-0.3 R_{L_{1}}$ in radius. We also computed $\beta$ values from
the observed $T_{{\rm b}}(r)$ distributions of other dwarf novae for
comparison. For Z Cha we find $\beta \simeq 0.2$ in quiescence (Wood
1990) and $\beta \simeq 0.8$ in outburst (Horne \& Cook 1985). For OY
Car we find values of $< 0.1$ and $\sim 0.7$, respectively for
quiescence (Wood 1990) and outburst (Rutten et al.  1992b). For V4140
Sgr we obtain $\beta = 0.36 \pm 0.02$ for the quiescent state and for
the decline from outburst. From the radial brightness distributions in
outburst we obtain $\beta = 0.25 \pm 0.03$.

Comparing these values of $\beta$, we see that the radial brightness
distribution of Z Cha and OY Car is flat in quiescence and closely
follows the $T \propto r^{-\frac{3}{4}}$ law of optically thick
steady-state disc in outburst (Horne \& Cook 1985; Wood et al.  1986;
Wood et al.  1989; Wood 1990; Rutten et al. 1992b).  In contrast, the
radial temperature distribution of V4140 Sgr is clearly steeper than
those of OY Car and Z Cha in quiescence, and becomes much flatter than
that of those stars in outburst. In Z Cha and OY Car the outbursts
occur with a significant increase in the relative brightness of the
inner disc regions, whereas in V4140 Sgr the outbursts reflect the
increase in brightness of only the outer disc regions. The small
amplitude and the flat temperature distribution of V4140 Sgr during
outburst may indicate a different physical origin when compared with
``normal'' dwarf novae outburst observed e.g. in Z Cha, OY Car or SS
Cyg.

Our distance estimate relies on the arguable assumption that the
quiescent disc of V4140 Sgr is optically thick. The inferred
disc temperatures depend on the assumed distance. We investigated the
effect of the distance on the derived brightness distributions, trying
to determine at which distance the binary should be in order to make
the brightness temperatures in the outbursting outer disc regions
higher than $T_{{\rm crit}}$.

\begin{figure}
\centering
\resizebox{\hsize}{!}{\includegraphics{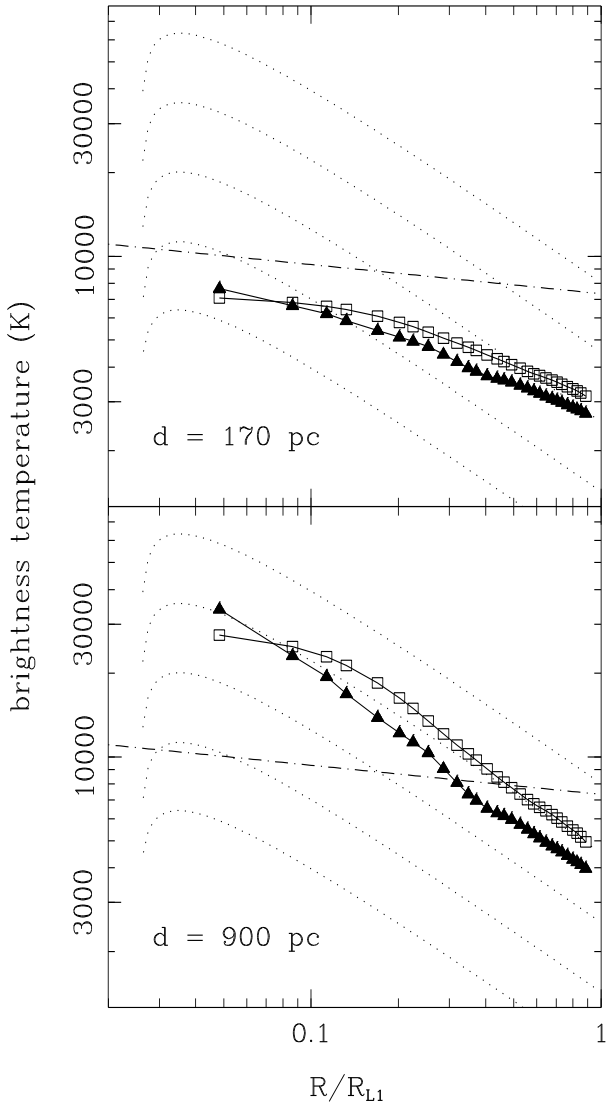}}
\caption{Average radial brightness temperature distributions for the 
  $B$ disc maps in outburst and quiescent states calculated for
  distances of 170 pc (top panel) and 900 pc (lower panel). For each
  panel, the points represents the mean temperature values computed in
  radius bins of $0.03 R_{L_{1}}$. Solid triangle and open squares
  correspond to the quiescence and outburst states (\emph{Bquies} and
  \emph{Bout}), respectively. Dotted lines correspond to steady-state
  optically thick disc models for mass accretion rates of $10^{-8}$,
  $10^{-9}$, $10^{-10}$, $10^{-11}$ and $10^{-12} M_{\odot}$yr$^{-1}$,
  from top to bottom respectively. A dot-dashed line shows the
  critical effective temperature $T_{{\rm crit}}$ below which the disc
  gas should remain to allow the thermal instability, and above which
  the gas should stay while in outburst (e.g. Warner 1995).}
\label{fig:distr_900}
\end{figure}

This only occurs if the distance is made larger than about $\sim 800$
pc. However, at this large distance the radial temperature
distribution in quiescence becomes steeper and closely follows the $T
\propto r^{-\frac{3}{4}}$ law, implying that the quiescent disc of
V4140 Sgr should be in a high viscosity, steady-state regime.  In
other words, the only way to remove the inconsistency regarding the
lower-than-expected temperatures in the outer disc is by making the
dwarf nova so bright that it becomes a nova-like variable.
Fig.~\ref{fig:distr_900} illustrate this by comparing the derived
radial temperature distributions in quiescence and outburst for
distances of 170 pc and 900 pc.

Thus, this exercise indicates that the small outburst seen in the
light curve of V4140 Sgr are very likely not caused by disc
instabilities according to the DI model. An alternative, plausible
explanation is that the outbursts of V4140 Sgr are episodes of
increased mass transfer from the companion star onto an accretion disc
permanently in a high viscosity regime. In this scenario, the low
amplitude of the outbursts of V4140 Sgr would be a consequence of its
quiescent, high-viscosity inner disc being hotter and brighter than
those of ``normal'' dwarf novae of similar orbital period (e.g. Z Cha
and OY Car). The comparison of eclipse shapes in quiescence supports
this idea. The pronounced orbital hump and the double-stepped eclipses
of Z Cha and OY Car reflect the fact that the quiescent disc is
relatively faint in comparison to the white dwarf and bright spot in
these stars.  On the other hand, the white dwarf, bright spot and
orbital hump features are much less conspicuous in the eclipses of
V4140 Sgr (Fig.~\ref{fig:obs_out}; see also Fig.~4 of BJS), indicating
that the quiescent disc is a relatively much more important source of
light than in Z Cha or OY Car.

V4140 Sgr may be a member of an increasing group of dwarf novae the
behavior of which cannot be explained by the DI model, together with
EX Dra (Baptista \& Catal\'{a}n 2001), EX Hya (Hellier et al 2000) and
V2051 Oph (Baptista \& Bortoletto 2004).

\section{Conclusions}
\label{sec:concl}

The main results of our photometric study of V4140 Sgr can be
summarized as follows:

\begin{itemize} 
  
\item[(i)] Our observations caught V4140 Sgr in the decline
  from an outburst in 1992 July and again in outburst in 2001 July.
  Further observations of a superoutburst of V4140 Sgr in
  2004 by amateur astronomers confirms its SU UMa-type dwarf
  nova classification. The observed outburst duration is $5-10$ days
  and the time interval between outbursts is $80-90$ days.
  
\item[(ii)]The observed amplitudes of the superoutbursts ($\Delta$mag
  $\simeq 2$) and of the regular outbursts ($\Delta$mag $\simeq 1$) of
  V4140 Sgr are remarkably low in comparison with the typical
  values for dwarf novae stars. Using the observed outburst amplitude
  and recurrence time interval, the object lies far away from the
  expected Kukarkin-Parenago relation.

\item[(iii)] The binary parameters were revised. The new primary mass
  $M_{1}$ is consistent with the observed mean white dwarf mass in
  CVs, and the new secondary mass $M_{2}$ is comparable to that of
  main sequence stars of same mass and radius.
  
\item[(iv)] A color-magnitude diagram was used to estimate the
  distance to V4140 Sgr, with a method similar to that used
  to constrain the distance to open clusters. Assuming an optically
  thick emission for the disc, a distance of $170 \pm 30$~pc is
  obtained.

\item[(v)] The radial brightness temperature distribution range from
  $3\,000$ K in the outer disc regions ($r \sim 0.4 R_{L_{1}}$) to
  $\sim 6\,000$ K in the inner disc regions ($r \sim 0.1 R_{L_{1}}$).
  The temperature distributions, both in quiescence and in outburst,
  are flatter than the $T \propto r^{-\frac{3}{4}}$ law expected for
  an optically thick steady-state accretion disc.
  
\item[(vi)] The derived disc temperatures are systematically lower
  than the critical temperature $T_{{\rm crit}}$ below which the disc
  gas should remain to allow outburst, both in quiescence and in
  outburst. The small outburst amplitudes and the flat temperature
  distributions may indicate that the outbursts are likely not caused
  by disc instabilities according to the DI model.
  
\item[(vii)] V4140 Sgr shows a peculiar behavior when
  compared to those of other dwarf novae with similar short orbital
  period. We suggest that the cause of the small amplitude outbursts
  of V4140 Sgr are episodes of enhanced mass transfer rate
  from the secondary star onto a disc in a permanent high-viscosity
  regime.

\end{itemize}

\begin{acknowledgements}
  We thank Christian Knigge for calling our attention to the error in
  the determination of the binary parameters of V4140 Sgr and
  an anonymous referee for useful comments that improved the
  presentation of the paper. In this research we have used, and
  acknowledge with thanks, data from VSNET, which are based on
  observations collected by variable star observers worldwide. This
  work was partially supported by CNPq/Brazil through research grant
  62.0053/01-1 -- PADCT III/Mil\^{e}nio. BWB acknowledges financial
  support from CAPES/Brazil. RB acknowledges financial support from
  CNPq through grant 300.354/96-7.
\end{acknowledgements}

\end{document}